\documentclass{acm_proc_article-sp} %
\pdfoutput=1               %
\usepackage{graphicx}      %

\usepackage{amssymb}
\usepackage{url}

\usepackage{amsmath}
\DeclareMathOperator{\face}{face}
\DeclareMathOperator{\spoof}{spoof}
\DeclareMathOperator{\blend}{blend}
\DeclareMathOperator{\cover}{cover}

\usepackage{rotating}    %
\usepackage{color}
\usepackage{etoolbox}    %

\usepackage{ulem}         %
\usepackage{booktabs}  %
\usepackage{comment}

\usepackage{xspace}    %

\usepackage[linesnumbered,ruled, vlined]{algorithm2e}
\SetKwProg{Fn}{Function}{\string:}{}
\SetKwInOut{Input}{Input}
\SetKwInOut{Output}{Output}

\newtoggle{comments}     %
\settoggle{comments}{true}
\iftoggle{comments}{
   \newcommand{\full}[1]{{\color{gray}#1}}
   \newcommand{\tocite}[1]{{\footnotesize\color{blue}[#1]}}
   \newcommand{\mjm}[1]{{\scriptsize\color{blue}[mjm: #1]}}
   
   \newcommand{\daw}[1]{{\footnotesize\color{magenta}[daw: #1]}}
}{
   \newcommand{\full}[1]{}
   \newcommand{\tocite}[1]{}
   \newcommand{\mjm}[1]{}
   \newcommand{\daw}[1]{}
}

\def\physicalchan/{{physical channel}}
\def\analogchan/{analog channel}

\usepackage
{hyperref}

\usepackage[all]{hypcap} %

\usepackage{refstyle}     %

\title{Spoofing 2D Face Detection: \\
Machines See People Who Aren't There}
\author{Michael McCoyd and David Wagner\\
  University of California, Berkeley
}
\date{}							%

\begin{document}
\maketitle

\begin{abstract}
Machine learning is increasingly used to make sense of the physical world
 yet may suffer from adversarial manipulation.
We examine the Viola-Jones 2D face detection algorithm to study whether 
  images can be created that humans do not notice as faces
  yet the algorithm detects as faces.
We show that it is possible to construct images that Viola-Jones
  recognizes as containing faces yet no human would consider a face.
Moreover, we show that it is possible to construct images that fool
  facial detection even when they are printed and then photographed.
\end{abstract}

\section{Introduction}
Machine learning is increasingly used to make sense of data from
sensors in the environment and for automated decision-making.
In this paper we look at 2D face detection.
For instance, face detection and face recognition have been used for
user authentication, tagging social media photos, video surveillance,
physical security, and other biometric security measures.

Similar to other biometrics, the security of 2D face detection and recognition 
   depends on whether it is used in attended or unattended settings.
For example,
  a door thumbprint reader in an empty corridor is vulnerable to attacks 
  that would not work in front of a guard at an entry gate.
Attacks on unattended use of facial biometrics have been well studied
from changing image perspective~\cite{duc_your_2009}
to using a 2D picture of an authorized person
 with cutouts for the eyes\cite{DBLP:conf/prisms/BoehmCFHKLMS13}.
However, these attacks might not be effective if attempted
   in front of guards or even casual bystanders, as such an attack is
   noticeable and easily detected.

The security of facial recognition in the attended setting has
not been well-studied.
In some applications, it is the attended setting that is arguably more
relevant.
For instance, facial detection and recognition might be
used for physical security and area access control;
because deployments might include the presence of a guard 
  or periodic video review, it is important to know whether there
  are attacks even a human wouldn't detect.
Facial recognition could also be used for authenticating to
a computer or end-user device (machine access control);
because there might be other employees present in the vicinity
who might have an opportunity to notice attacks,
the attended setting is relevant here as well.
In either situation, holding up pictures of authorized individuals could raise alarms.
In this paper, we study attacks on facial detection in the
attended setting: we study whether it is possible
to fool face detection algorithms
  into thinking an extra face is present,
  while preventing humans (e.g., security guards, others in the vicinity)
  from noticing the extra face.

\begin{figure}[tbp]
\begin{center}
\includegraphics[width=.333\columnwidth]{%
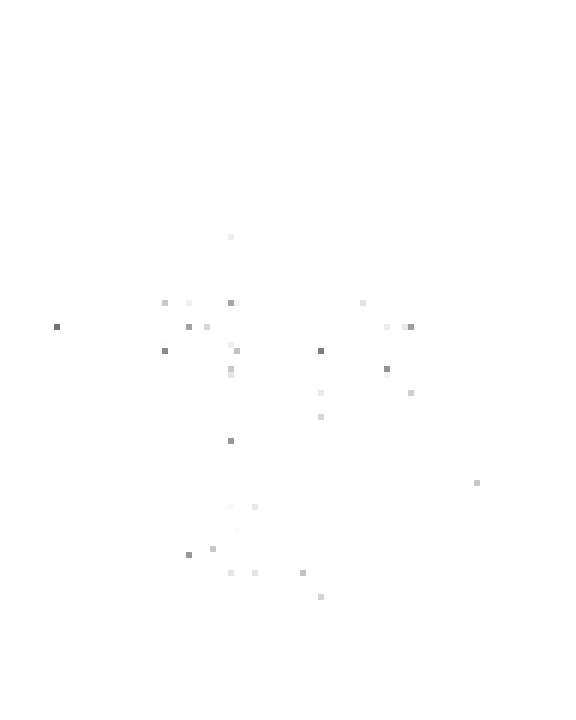}
\includegraphics[width=.333\columnwidth]{%
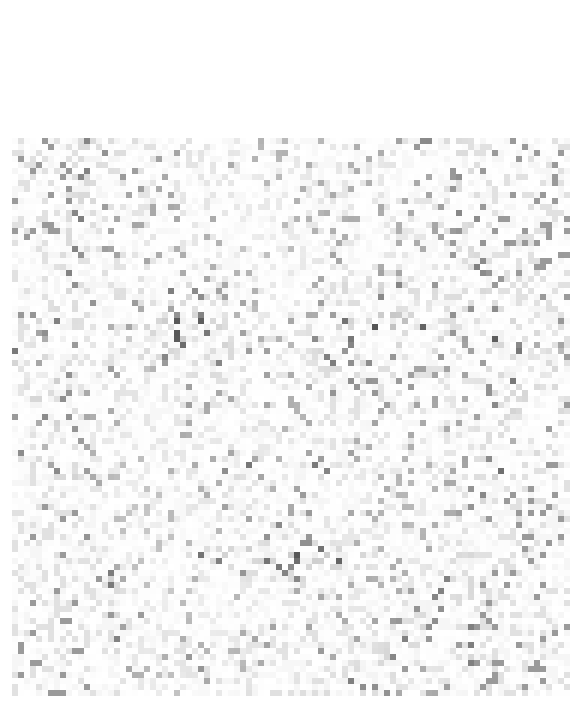}
\caption{Two malicious images we generated with our algorithms.
The left image is detected as a face, if fed directly into a face detector.
The right image is detected as a face, even after it is subjected to distortions
imposed by the physical world: when displayed on a tablet and
captured using a camera, the resulting image is detected as a face
97\% of the time.
}
\figlabel{fig:successful_attacks}
\end{center}
\end{figure}

Why study facial detection, rather than facial recognition?
Ultimately, it is the ability to fool facial recognition that matters.
However, facial detection algorithms are simpler to study.
In this paper we focus on the security properties of facial detection.
We view this as a first step towards the longer-term goal of analyzing
facial recognition.
Also, because facial recognition algorithms typically begin by first using
facial detection to look for faces, then apply facial recognition to
each detected face, any attack on facial recognition must begin by
first fooling the facial detection step.
Thus, we believe our techniques may have lasting relevance.

The most commonly used algorithm for facial detection,
Viola-Jones\cite{Viola01robustreal-time} \secref*{sec:VJ},
works using machine learning techniques to build a classifier
that determines whether a region of an image contains a face or not.
We construct successful attacks on the Viola-Jones algorithm.
One technical challenge is that the facial detection algorithm only outputs
a binary signal: face or non-face.
Thus we can not use normal techniques for fooling classifiers,
   such as attempting to solve for a solution 
   or using gradient descent to search for inputs that fool the classifier.
Thus, we are forced to devise novel algorithms for constructing
images that will fool the Viola-Jones classifier.

A second challenge is that the attacker cannot perfectly
control the input to the classifier.
Unlike spam, the adversary does not directly control the input to the
classifier.
Rather, the image passes through a noisy \physicalchan/ 
  -- the adversary can display one image, but the image captured
  by a camera will be reproduced only imperfectly.
The signal is degraded by blurring, random noise, and other effects.
We devise attacks that are robust to these effects.

Our approach uses a feedback-guided search algorithm to construct
an image that Viola-Jones recognizes as a face, yet is unlikely to be
recognized by a human.
We select a cover image $C$ that does not contain a face; for instance,
$C$ might be simply an all-white image.
We start with an ordinary image of a face (recognized as a face by
humans and Viola-Jones alike), $F$, and iteratively modify $F$.
At each step we make a small random modification to $F$ to make it
more similar to $C$, but while ensuring that $F$ remains recognized as
a face by Viola-Jones.

Essentially, our algorithm uses the Viola-Jones classifier to provide
feedback and guide a directed random walk through the space of images,
probing the decision boundary of the classifier to search for an image
that is as similar to $C$ as possible while still being classified as
a face by Viola-Jones.
Through appropriate instantiation of this approach, we are able
to create digital spoof images that humans do not notice
  yet Viola-Jones detects,
  if they are presented directly to the facial detection algorithm
  with no modifications.

We then refine our algorithm to deal with degradation during delivery
imposed by the physical world.
We imagine conducting our attack in a simplified physical world, 
  which we model with a simulated \analogchan/, \secref*{sec:phys_analog}.
By modeling the effects observed in our experiments, we are
able to create a reasonable simulation of the
kinds of degradation imposed by the physical world
and then adjust our attack (\secref*{sec:attack_analog})
to create spoof images that are more robustly detected
despite degradation imposed by the physical world.

Our attacks are successful.
See \figref{fig:successful_attacks} for two examples of
malicious images we generated using our techniques.

The contributions of this work are that we show new attacks
on facial detection; introduce a new algorithm to construct inputs
that fool a classifier, using binary outputs from the classifier;
and devise techniques for dealing with noise and image degradation
introduced by the physical world.
The next step for future research is to study the security of
facial recognition.
If it is possible to extend our results to spoof facial recognition as well,
this might enable new, stealthier attacks on facial recognition:
e.g., an attacker might be able to print a spoof image on the
  front cover of a notebook and casually hold in view of a security
  camera, thereby gaining access to a protected computer or physical
  location.
We leave analysis of this threat to future work.

\section{Background}\seclabel{sec:VJ}
The industry standard for face detection is the Viola-Jones classifier\cite{Viola01robustreal-time}.
It accepts a grayscale image and produces as output a boolean value,
indicating whether the image is a face or not.
Typically, to detect all faces in an image, we run the Viola-Jones classifier
over all regions of the image and see which ones it classifies as a face.

For completeness we provide a concise overview of Viola-Jones classifier,
but for purposes of this paper, it is not important to understand the
details of how the Viola-Jones classifier works.
It is a boosted cascade classifier built out of multiple weak classifiers
and trained on a set of face and non-face images.
Each weak classifier computes the average intensity within two rectangles,
subtracts these two numbers, and compares the result to a threshold
to decide whether 
 that portion of the image might contain a face (see \figref{fig:viola_jones}).
Early stages of the cascade use only a few weak classifiers with large rectangles;
   later stages use more classifiers with smaller rectangles.
As a result, the Viola-Jones algorithm is very fast.

\begin{figure}[tbp]\begin{center}
\includegraphics[width=.22\columnwidth]{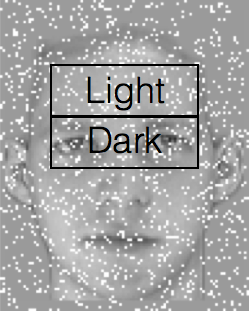}
\includegraphics[width=.22\columnwidth]{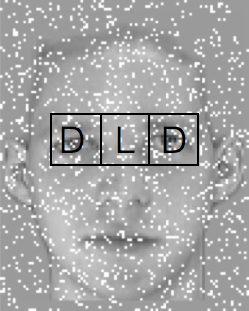}
\caption{Two of Viola-Jones' early weak classifiers.
}
\figlabel{fig:viola_jones}
\end{center}\end{figure}

\section{Problem Statement}

Our attack model in this work
 is that the adversary might carry anything stealthy that lets
 them display an image crafted to fool facial detection.
For example, one possible attack might be to carry a notebook
 with a printed image of a spoof image while approaching
  a security access point.
The attacker could casually hold the notebook against his/her
 chest and neck,
   as if they had just consulted some notes but were now finished with the notes,
   as they walk through the access point.
Our attacks aim to cause the automated security system to notice
one extra person.
Ultimately, the goal would be to extend our attacks to fool facial
recognition as well, so the system sees an authorized person who is not
actually there and allows the attacker through --- though we have not
studied facial recognition, so our work should be viewed as
only a first glimpse at what ultimately might be possible.

Many variants of this attack scenario are possible.
Instead of a printed image, the attacker could carry a flat-panel display 
   as part of the cover of a notebook, which might allow finer control over
   the displayed image --- this is the scenario we focus on in this paper.
One could also imagine an attacker who seeks to gain
  access to a computer that uses facial-recognition-based login
  (rather than password-based login).
An attacker might find a publicly available
  image of the face of an authorized user, then use our techniques to
  try to disguise that image.

\section{Starter Attacks}\seclabel{sec:starter}
Our attack starts with an ordinary image of a face,
and morphs it until it is no longer recognizable to humans as a face.
Roughly speaking, we do this by
blending in enough of a cover image so the original face is no longer
detectable to humans,
while preserving enough of the original face so that
a face detection algorithm%
\footnote{Viola-Jones as implemented by OpenCV v3.1.1 using the haarcascade\_frontalface\_default.xml template file with 
cv2.CascadeClassifier.detectMultiScale called with default parameters plus CASCADE\_SCALE\_IMAGE}
still detects the face.
 
\paragraph{Data}
To illustrate our attacks we use faces from the AT\&T Database of Faces\cite{ferdin94:285}.
Each face is a  8-bit $112  \times 92$ grayscale image with a dark background.
Before applying our attacks,
 we use Photoshop's magic wand tool to replace the dark background and hair 
  with a background the same tone as the face.
We also make the images 120 pixels tall by 96 wide by adding extra border pixels, to provide a bit more robustness to cropping.
See \figref{fig:face_data} for an example.

\begin{figure}[tbp]
\begin{center}
\includegraphics[width=.22\columnwidth]{%
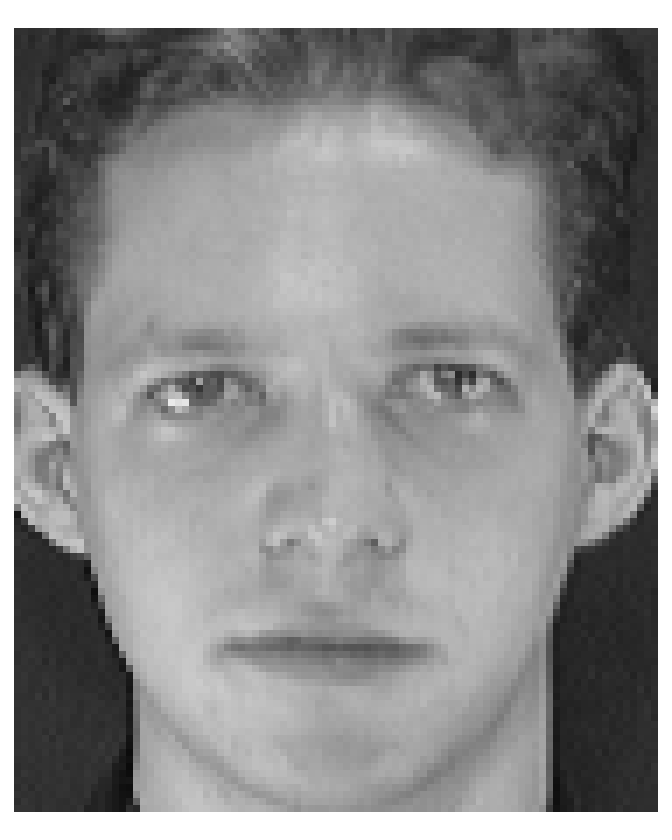}
\includegraphics[width=.22\columnwidth]{%
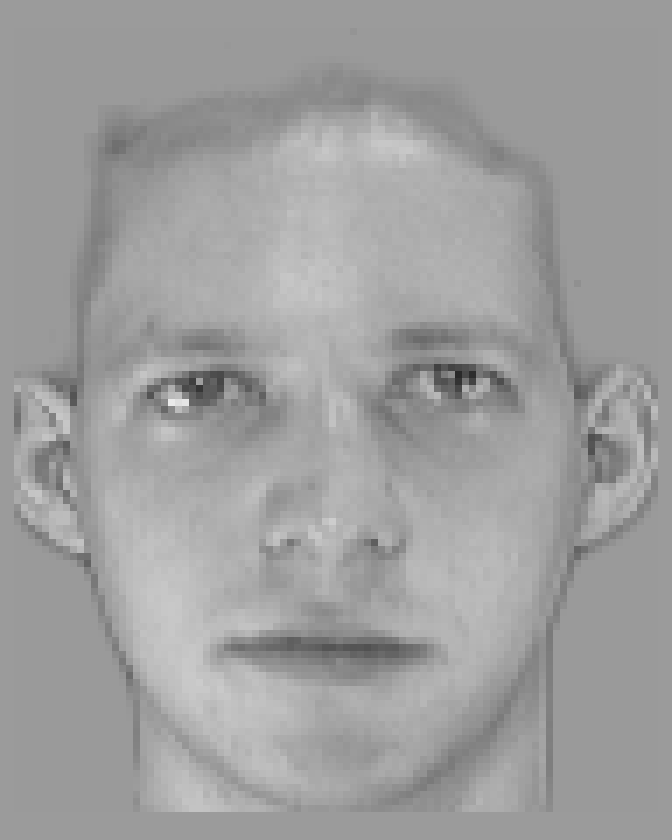}
\caption{An example face from the AT\&T database
(left), and adjusted with flesh-toned background (right).
}
\figlabel{fig:face_data}
\end{center}
\end{figure}

\subsection{Starter Attack: Blend}
A very simple attack would be to construct to the spoof image
as a blend  of the face and cover image.
Each pixel, $p$, receives a fixed percent, $r$, of the cover value, 
 with the remaining percentage from the face value.
Specifically,
$$
\spoof_r(p) = r \times \cover(p) + (1 - r) \times \face(p)\text{,}
$$
 where $r \in [0, 1]$  is constant for all $p$.
We increase $r$ until spoof$_r$ is no longer detected by Viola-Jones.
\Figref{fig:attack_blend} illustrates the attack with a cover image of granite.

This simple attack is unsuccessful.
With a cover image of granite, Viola-Jones detects a face
only for values of $r$ in the range $0$--$0.16$; humans can
recognize a face for any value of $r$ in the range $0$--$0.58$, so
there is no value of $r$ which is accepted by Viola-Jones but not
detected by humans.
See \figref{fig:attack_blend} for an illustration.

We tested three cover images and the attack fails for all three.
With cover images of 
 granite, sand, and all gray,
 values of $r$ above
   $16\%,  8\%, \text{and } 45\%$ fail Viola-Jones detection.
Much higher values are needed to start deceiving the human authors' detection,
  $58\%,  74\%, \text{and } 95\%$ respectively.
There is no overlap between the images where humans don't see a face and images where Viola-Jones detects a face.

\begin{figure}[tbp]
\begin{center}
\includegraphics[width=\columnwidth]{%
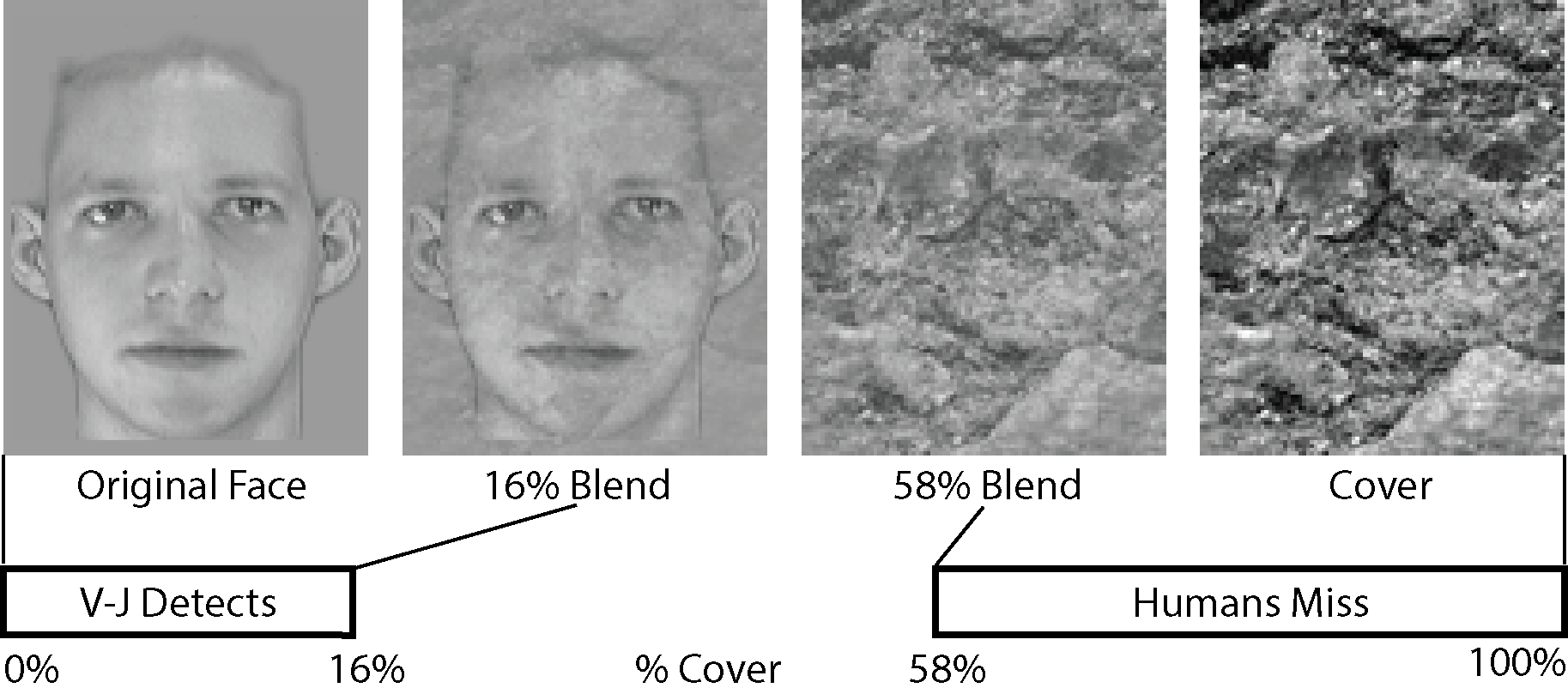} 
\caption{
Blend attack showing the original face, two blends, and the cover image.
The blends are the maximum percent of cover image that can be blended in
  while ensuring Viola-Jones still detects a face,
  and the minimum percent that yields something not recognized as a
  face by humans.
The attack fails as no blend percentages meet both conditions.
}
\figlabel{fig:attack_blend}
\end{center}
\end{figure}

\subsection{Starter Attack: Random Subset of Pixels}\seclabel{sec:starter_choice}

We also tried a different strategy: instead of blending all pixels,
pick a random subset of pixels and replace them entirely with the corresponding
pixel from the coverage image.
One might hope this will preserve the region differences used by Viola-Jones
  yet introduce enough detailed noise to fool the human. 
Specifically, for a given face, cover image, and fraction $r$, we choose
$$R = \ \text{random set of $r$ fraction of the pixels}$$
and then define the spoof image as
$$\spoof_r (p) =
  \begin{cases}
     \cover (p) & \text{if } p \in R \\
     \face(p)       & \text{otherwise}
    \end{cases}
$$

To test this approach, we created 300 random images for each
choice of four possible cover images and $r$ in the range $[0.00,1.00]$
in steps of $0.01$.
Cover images of granite, sand, all-gray, and all-white were used. 
The authors viewed five spoofs for each $r$, starting from $r = 1.00$, and judged when a face is first human-recognizable.

\begin{figure}[tbp]
\begin{center}
\includegraphics[width=\columnwidth]{%
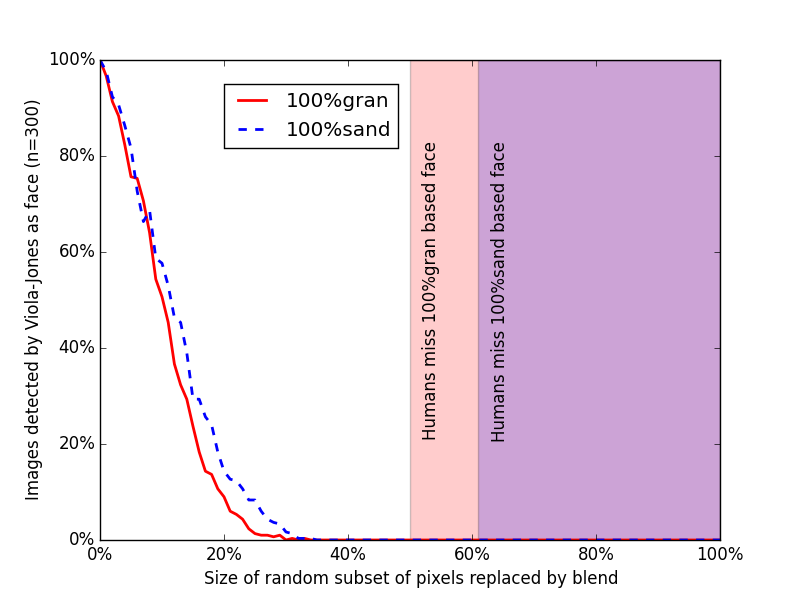} 
\caption{
Random subset of pixels attack.
The y-axis shows the fraction of candidate spoof images that were
detected by Viola-Jones as a face.
There are no images where the face is detected by Viola-Jones, yet missed by the humans. 
Only images built from granite or sand hid the face from the humans.
}
\figlabel{fig:attack_rand_choice}
\end{center}
\end{figure}

This attack also is unsuccessful.
Only the granite and sand cover images had results that were at all
promising.
However, as \Figref{fig:attack_rand_choice} shows, no spoofs were simultaneously detected as faces by Viola-Jones 
  and missed by the humans. Thus this attack fails.
But in the tail of the graph
  there are images that Viola-Jones detected with $r=29\%$, which is more than we saw in the blend attack, where $r=16\%$ was the maximum achievable.

\subsection{Starter Attack: Random of Blend}\seclabel{sec:starter_random_blend}
Another natural idea is to combine the previous two attacks.
A slightly better attack might 
   replace
  a percent, $r$, of random pixels from the original face
   with some blend of the face and cover image.
For example, using the images of \figref{fig:attack_blend}, 
we could replace some percent of random pixels in the original face
with the pixels from the 58\% blend of granite.
Specifically, for a given face, cover image, blend rate $b \in [0, 1]$, and fraction of blend pixels $r \in [0, 1]$, 
the spoof image is defined as
\begin{align}
\blend_b (p) =& \ b \times \cover(p) + (1 - b) \times \face(p)\text{,} \\
R =& \ \text{random set of $r$ fraction of pixels}. \\
\spoof_r (p) =& 
  \begin{cases}
     \blend_b (p) & \text{if } p \in R \\
     \face(p)       & \text{otherwise}
    \end{cases}
\end{align}

\begin{figure}[thbp]
\begin{center}
\includegraphics[width=\columnwidth]{%
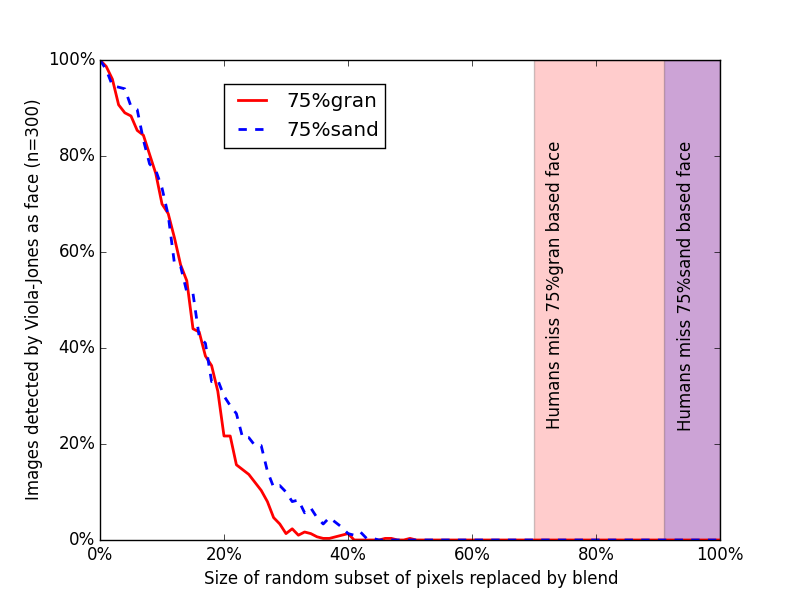} 
\caption{
Random single blend  attack:
Detection rate of faces that replace a random set of pixels with a blend of the face and cover images.
There are no images where the face is detected by Viola-Jones, yet missed by the humans. 
Only images built from 75\% blends of granite or sand hid the face from the humans.
All other images, including based on blends of gray or white, failed to hide the face from the humans. 
}
\figlabel{fig:attack_rand_blend}
\end{center}
\end{figure}

We evaluated this attack in the same manner as the random choice attack,
  except that
  for each cover image we created three blends of 25\%, 50\%, and 75\% cover.
A random 300 images were created for each combination of cover image, blend, and value of $r$.

Our results, in \figref{fig:attack_rand_blend},
 show that we again can not satisfy Viola-Jones while fooling humans.
  Viola-Jones detection rates drop as the blend rate and random set size increase,
  and only two of the blends create a small range of images that hide the face from the authors. 
However,
 the bottom tails of the curves continue to show there are some series of choices of pixels 
  that continue to succeed even after most other series of choices of pixels long ago failed. 
This suggests that perhaps further improvement might be possible by adding some guidance from Viola-Jones about the shift amount and/or choice of pixels.

\section{Attack: Random Shift (Exact)}\seclabel{sec:attack_exact}
We build on these ideas to construct a more sophisticated attack.
The attack applies small random changes to the image but
reverts any changes that cause Viola-Jones to fail to detect a face.
From our starter attacks, we learned that 
  changing pixels only part way to the cover image allows us to retain Viola-Jones detection longer,
  and that some sets of pixel choices work better than others.
We combine these lessons by
 shifting random pixels small amounts toward the cover image 
 and undoing those changes that cause a failure.

Our attack procedure has two parts:
  a search routine picks a suitable attack image while
  an oracle evaluates that attack image.
The search is very simple.
We have a loop that 
  picks a random pixel, 
  changes its intensity halfway closer to the intensity of the corresponding pixel in the cover image,
  but rejects the change if the oracle says the face is no longer detected.
The oracle checks whether a face would be detected in the spoof image
  at the same location as it is detected in the original face image.
We first develop our attack, algorithm~\ref{alg:exact},
  with an oracle that passes the attack image directly to Viola-Jones,
  without accounting for any degradation that might be caused by
  the physical channel.

\begin{algorithm}  %

\Fn{Search($F$, $C$)}{
    \Input{Face image $F$,\\
              Cover image $C$} 
    \Output{Spoof image $S$}

$S \leftarrow F$\;
\While{not Stalled($S$)}{
   $p' \gets $ a random pixel in $S$   %
   
  $\displaystyle
    T(p) \leftarrow 
  \begin{cases}
     \text{Round} \left( \dfrac{S(p) + C(p)}{2}  \right) & \text{if } p = p' \\
     S(p)       & \text{otherwise}
    \end{cases}
  $

  \If{Oracle($T$)}{
     $S \leftarrow T$\:
  }
  } %
return S
} %

\Fn{Stalled($S$)}{
    \Input{Spoof Image $S$} 
    \Output{Boolean}
    Return True if $S$ has not changed in last several calls.
}

\Fn{Oracle($T$)}{
    \Input{Test Image $T$} 
    \Output{Boolean}
    Return true if Viola-Jones detects a face in $T$ with bounds within $10\%$ of those of the face in $F$.
}

\caption{Exact attack}
\label{alg:exact}
\end{algorithm}   %

\paragraph{Experiment}
In preparing the data,
   we replace the area around the face with the corresponding pixels from the cover image
      (see \figref{fig:exact_inputs}).
Much of the image background is not used by Viola-Jones,
   when detecting the face centered in the attack image,
  so we use Viola-Jones to detect the location of this central face
  and preprocess our starting face to retain just the face area pixels.

\begin{figure}[tpb]
\begin{center}
\includegraphics[width=.22\columnwidth]{%
figs/00_StartFace20160629/big7_s1_1alphaBorder120}
\includegraphics[width=.22\columnwidth]{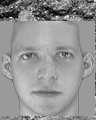}
\caption{The face we used for the exact attack:
the original AT\&T image (left) and
after altering the background and pasting the coverage image
around the face area (right).
}
\figlabel{fig:exact_inputs}
\end{center}
\end{figure}

\paragraph{Results}
This attack is successful.
Because of our algorithm, all the results of our exact algorithm pass Viola-Jones detection.
Typical results can be seen in the top row of \figref{fig:exact_results},
  for cover images of granite, sand, all-white, and all-gray.
The granite, sand, and white attack images do not seem to stand out as faces to the authors,
 with the white attack image being particularly nice.
In the other spoofs we can see a grid pattern in the face region,
  presumably an artifact of the Viola-Jones regions.

\begin{figure}[tpb]
\begin{center}
\begin{tabular}{c@{\hskip3pt}c@{\hskip1pt}c@{\hskip1pt}c@{\hskip1pt}c@{\hskip1pt}c}
\begin{turn}{90}$\quad$ Spoofs\end{turn}&
\includegraphics[width=.22\columnwidth]{%
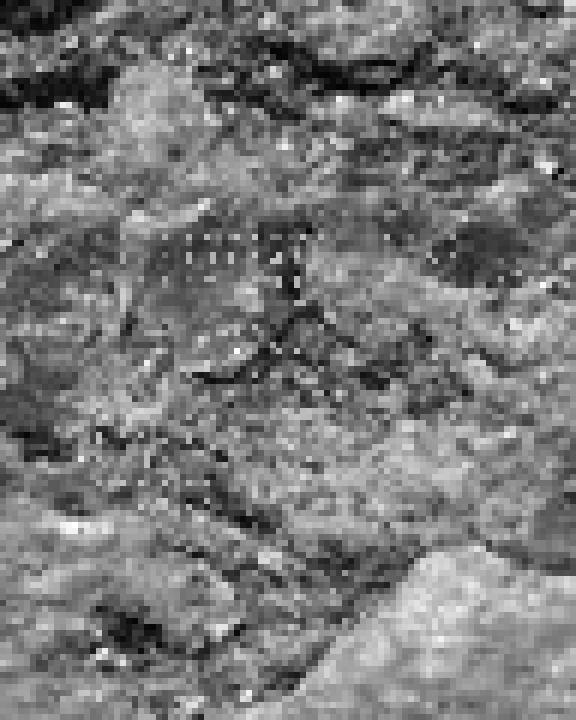}&
\includegraphics[width=.22\columnwidth]{%
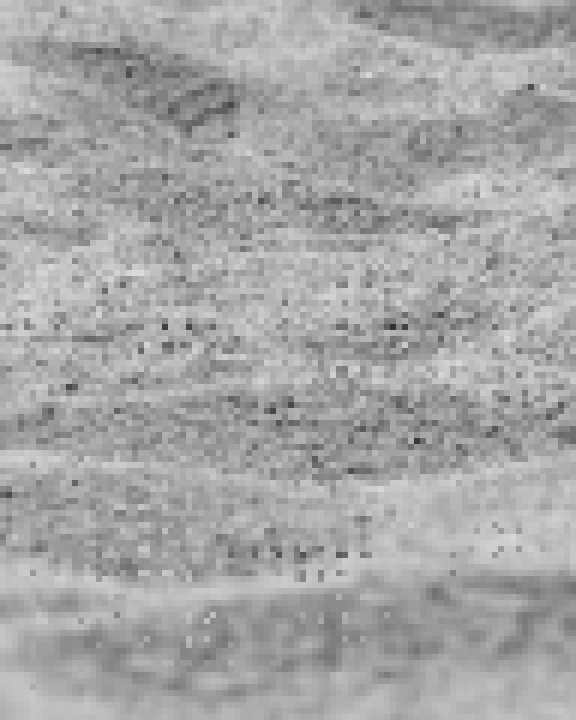}&
\includegraphics[width=.22\columnwidth]{%
figs/05_20160713_rand_exact/white/big_spoof_0p5_S____63,121_Y}&
\includegraphics[width=.22\columnwidth]{%
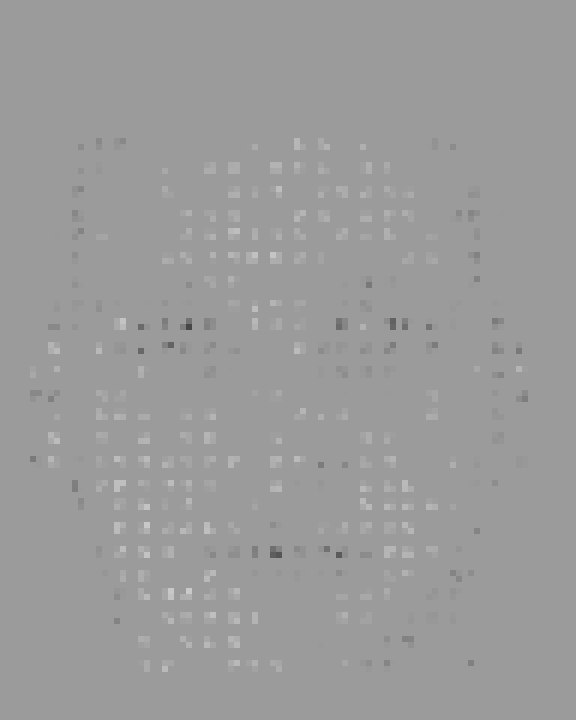}\\
\begin{turn}{90}$\quad$After Physical\end{turn}&
\includegraphics[width=.22\columnwidth]{%
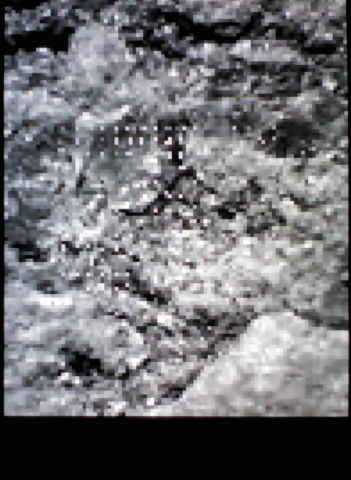}&
\includegraphics[width=.22\columnwidth]{%
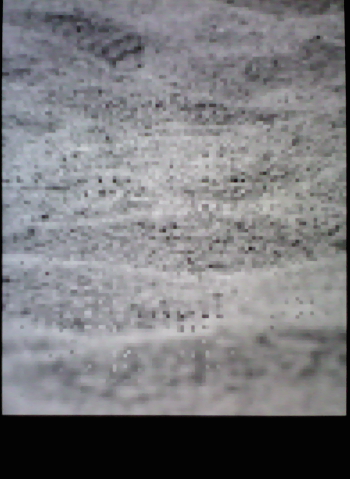}&
\includegraphics[width=.22\columnwidth]{%
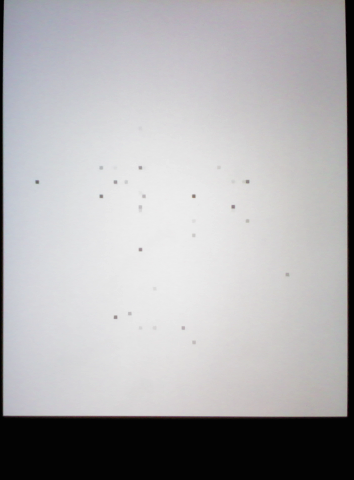}&
\includegraphics[width=.22\columnwidth]{%
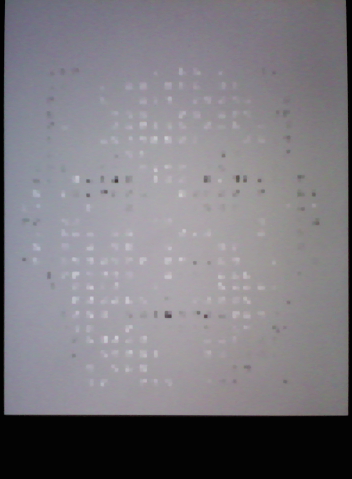}\\
\end{tabular}
\caption{Results of the exact attack.
The spoof images are detected as faces if fed directly into Viola-Jones.
However, they are not detected by Viola-Jones after passing through the physical world.
}
\figlabel{fig:exact_results}
\end{center}
\end{figure}

\paragraph{Discussion}
The attack images in the results are successful in terms of creating images
    detected by Viola-Jones as a face but
    not noticed by humans as faces.
Yet there is a problem. 
They are not detected as faces if we 
  display these attack images on a retina resolution tablet,
  view them through a webcam as in \figref{fig:physical_world},
  and pass the resulting webcam capture to Viola-Jones.
The changes present in the resulting images can be seen in the bottom row 
  of \figref{fig:exact_results}.
The failure of our spoof images in the physical world motivates the next steps of our attack.

\section{Physical and Analog Channels}\seclabel{sec:phys_analog}
To extend our attack to the physical world,
  we created a simplified physical world and a simulation of it, an \analogchan/.
We will use the term \textit{\physicalchan/} for passing an image through our display and webcam in a box,
   and the term \textit{\analogchan/} for our software simulation of that.

\paragraph{Physical Channel}
To build a reproducible physical attack environment, we display the image on a tablet with a retina display in a low glare box with a fixed webcam. See \figref{fig:physical_world}.
\begin{figure}[tbp]\begin{center}
\includegraphics[width=.15\textwidth]{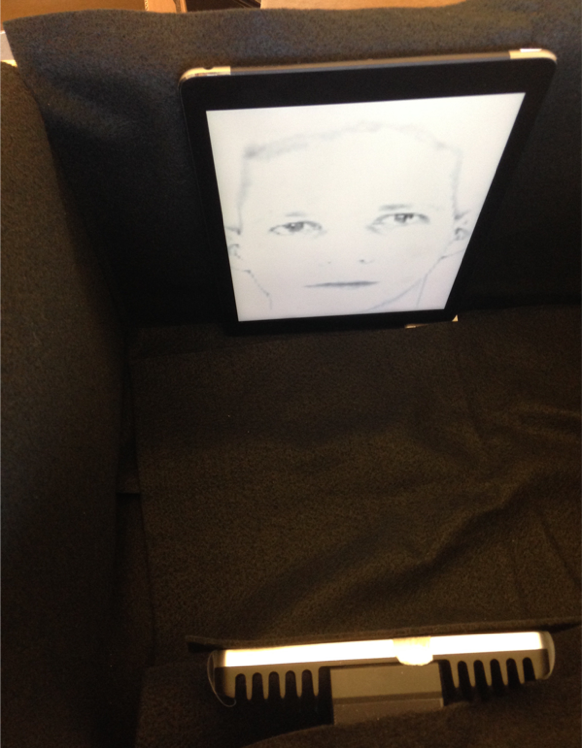}
\caption{Our simplified physical world has a tablet with retina and a fixed webcam in a low-glare box.}
\figlabel{fig:physical_world}
\end{center}\end{figure}
To reduce the effect of browser image smoothing, we expand the image several times before display, 
  resulting in image pixels displayed as crisp squares.
Apart from the fact that the display position is fixed, this is a fairly realistic simulation of an attack scenario, as the attacker can carry a no-glare tablet
  and security webcams typically have fixed locations.
We then measured the effect of the physical channel on images.

\paragraph{Individual Effects}
We found that passing images through our physical world has seven effects.
We model five of them:
  brightening the image center,
  adding noise,
  adding Gaussian blur,
  reducing dark contrasts,
  and replicating pixels.
We do not model barrel distortion, as we assume its effect is slight.
We also do not yet model differences in alignment between the image and camera pixel borders. 
We created test images to help us measure each of these effects.
As we consider each effect, in each figure we show on the left a test image, in the center an image after the \physicalchan/, and on the right an image after our \analogchan/.

To measure center brightening, we displayed a uniformly gray test image and captured 300 video frames of it.
Averaging these frames gives us the average intensity for each pixel, \figref{fig:effects_center}, 
that the camera sees when this uniform gray image is displayed.
We simulate this effect by 
 adding to the input image the difference between the average image and the average image's average pixel value.
Our analog channel winds up a bit darker than the physical channel.

\begin{figure}[tbp]\begin{center}
\begin{tabular}{c@{\hskip1pt}c@{\hskip1pt}c}
\includegraphics[width=.15\textwidth]{%
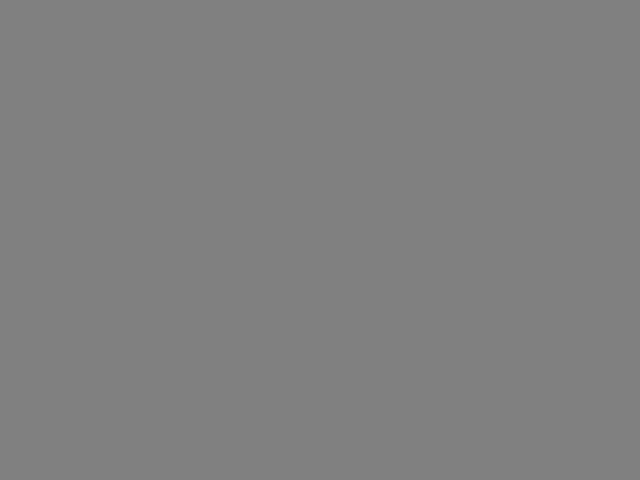} &
\includegraphics[width=.15\textwidth]{%
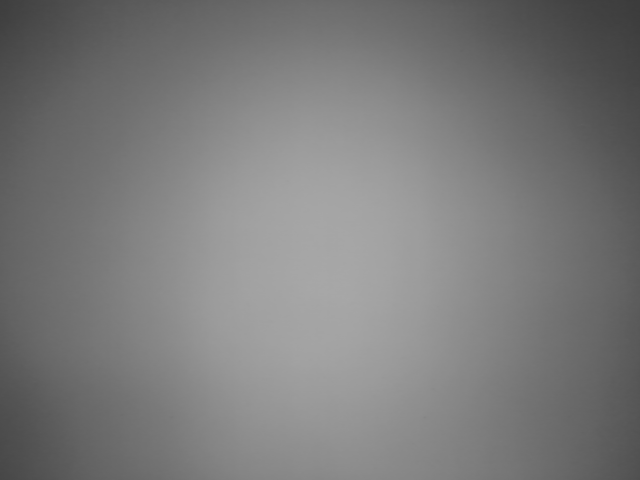} &
\includegraphics[width=.15\textwidth]{%
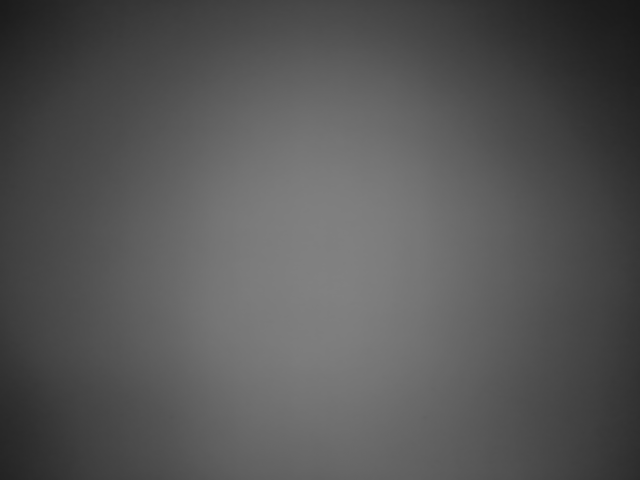}\\
Test Image & After Physical  & After Analog \\
\end{tabular}
\caption{Channel center brightening.
}
\figlabel{fig:effects_center}
\end{center}\end{figure}

The channel noise can be seen by first removing the center brightening effect.
We subtract the average image from a single frame, and then add back the test image.
A detail of the result can be seen in \figref{fig:effects_noise}.
We get a noise distribution by subtracting from each frame the average image
and estimate the distribution of the resulting differences (across all pixels).
We found that this distribution is well-fit by a normal distribution
with mean zero and standard deviation $1.5$ (see \figref{fig:effects_noise_stats}).
Thus, to approximate the noise, our analog channel adds Gaussian noise
with standard deviation $1.5$ to the intensity of each pixel.

\begin{figure}[tbp]\begin{center}
\begin{tabular}{c@{\hskip3pt}c@{\hskip1pt}c@{\hskip1pt}c}
\begin{turn}{90}$\quad$ Original\end{turn}&
\includegraphics[width=.12\textwidth]{%
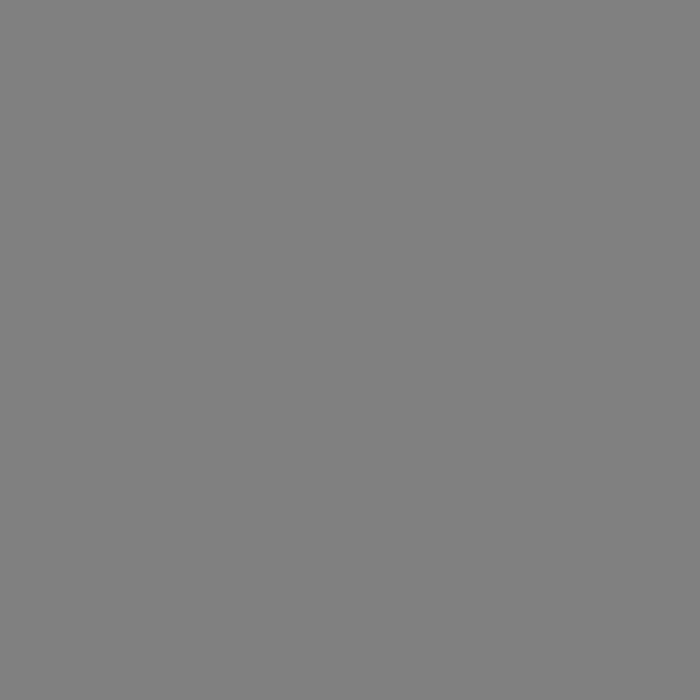}&
\includegraphics[width=.12\textwidth]{%
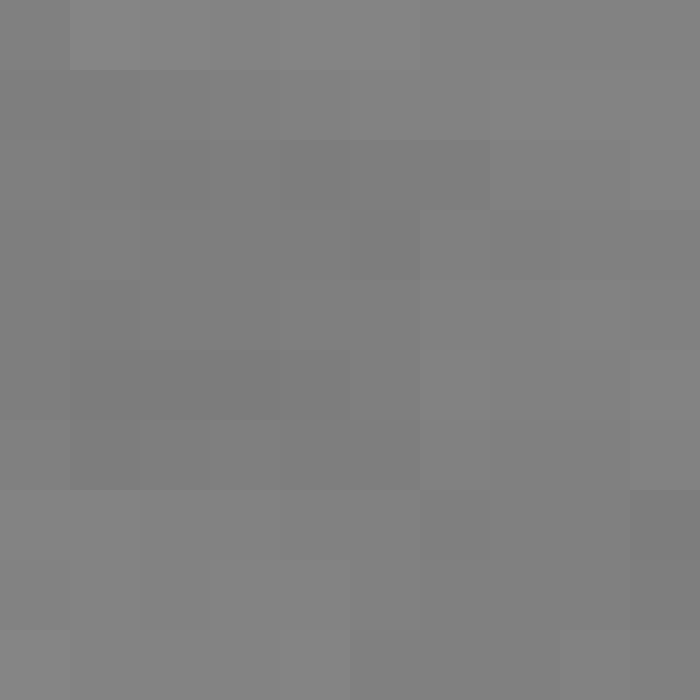}&
\includegraphics[width=.12\textwidth]{%
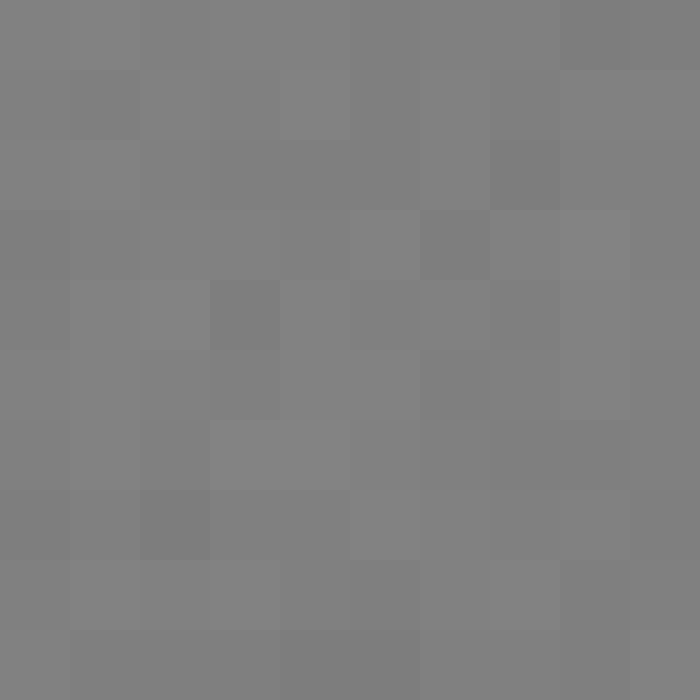}\\
\begin{turn}{90}$\quad$ Enhanced\end{turn}&
\includegraphics[width=.12\textwidth]{%
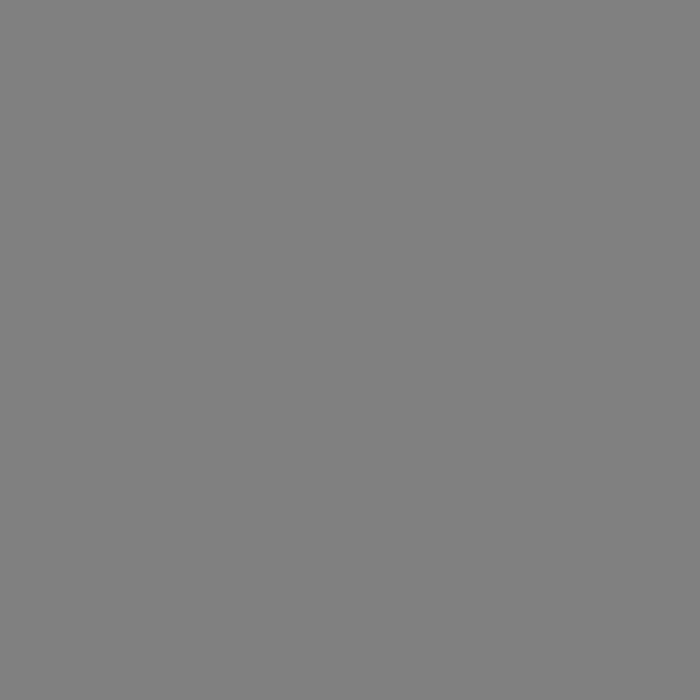}&
\includegraphics[width=.12\textwidth]{%
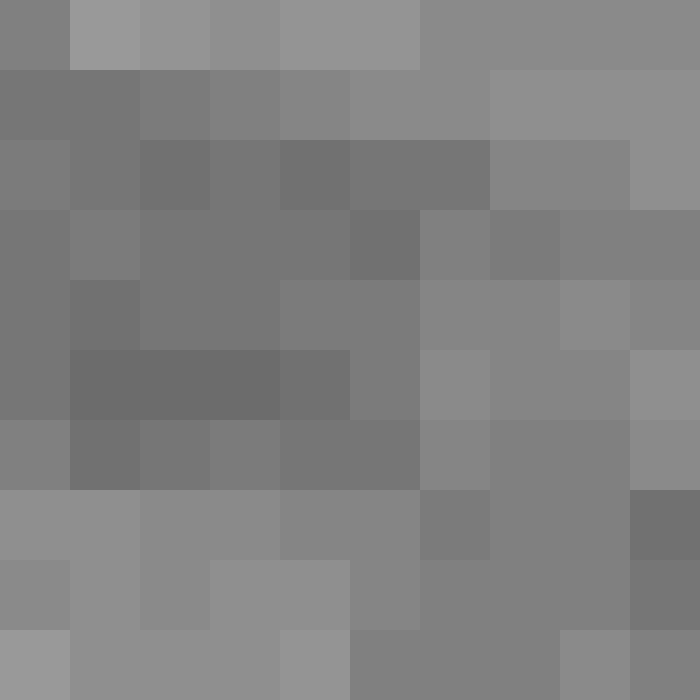}&
\includegraphics[width=.12\textwidth]{%
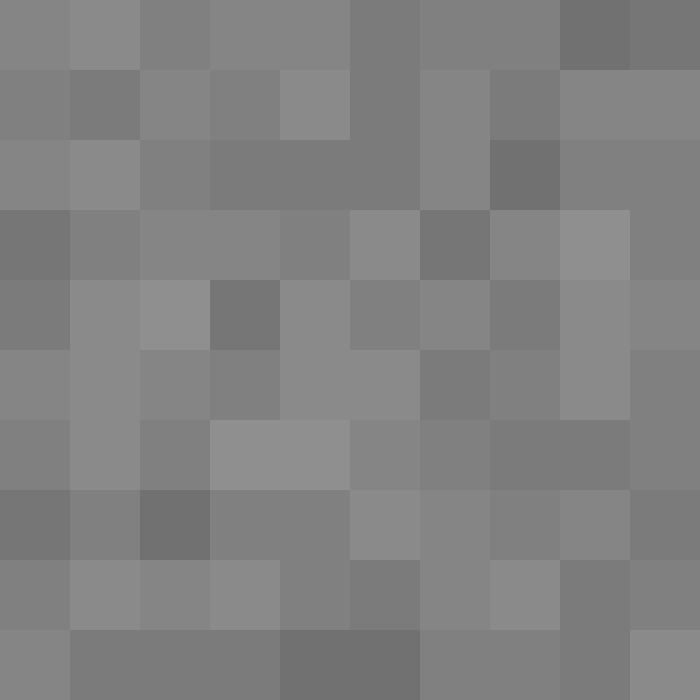}\\
& Test Image & After Physical  & After Analog \\
\end{tabular}
\caption{An example of channel noise seen in a 10 by 10 pixel detail of a gray image. The bottom row enhances the intensity differences by five.
}
\figlabel{fig:effects_noise}
\end{center}\end{figure}

\begin{figure}[tbp]\begin{center}
\includegraphics[width=.45\textwidth]{%
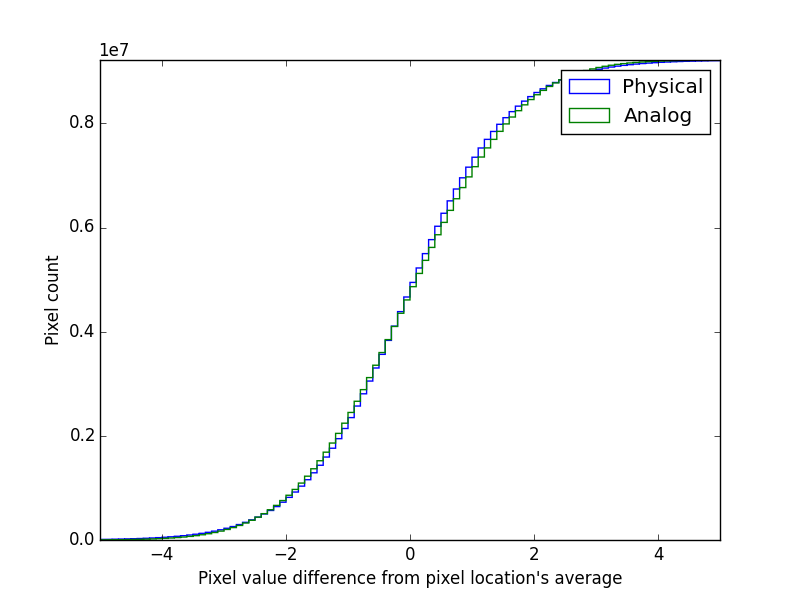}
\caption{The distribution of additive noise imposed by the channel to each pixel intensity: we show both the observed distribution from the physical channel and the distribution of noise added by our simulated analog channel.
}
\figlabel{fig:effects_noise_stats}
\end{center}\end{figure}

To assess channel blur, we displayed a test image with vertical bars whose width is one to four pixels.
The result can seen in \figref{fig:effects_blur}.
Our analog channel uses a Gaussian blur with standard deviation $0.9$,
as we found that this matches the observed effects well.

The physical channel has a non-linear effect on pixel intensities:
for example, it reduces the contrast in
dark regions and increases the contrast for light intensities, as shown
in \figref{fig:effects_contrast}.
We found that this can be modelled with a piecewise-linear intensity
response curve applied uniformly to all pixels: a pixel with intensity
$x$ becomes intensity $f(x)$, where $f$ is a piecewise-linear function.
We found that two pieces are sufficient to fit the observed response
curve well.

\begin{figure}[tbp]\begin{center}
\begin{tabular}{c@{\hskip3pt}c@{\hskip1pt}c@{\hskip1pt}c}
\includegraphics[width=.15\textwidth]{%
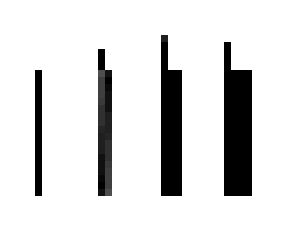}&
\includegraphics[width=.15\textwidth]{%
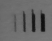}&
\includegraphics[width=.15\textwidth]{%
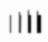}\\
Test Image & After Physical & After Analog\\
\end{tabular}
\caption{Channel blur effects.
}
\figlabel{fig:effects_blur}
\end{center}\end{figure}

\begin{figure}[tbp]\begin{center}
\begin{tabular}{c@{\hskip3pt}c@{\hskip3pt}c@{\hskip3pt}c}
\includegraphics[height=.2\textheight]{%
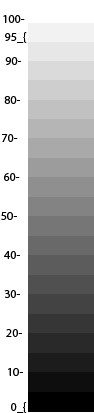}&
\includegraphics[height=.2\textheight]{%
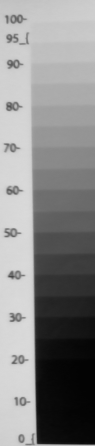}&
\includegraphics[height=.2\textheight]{%
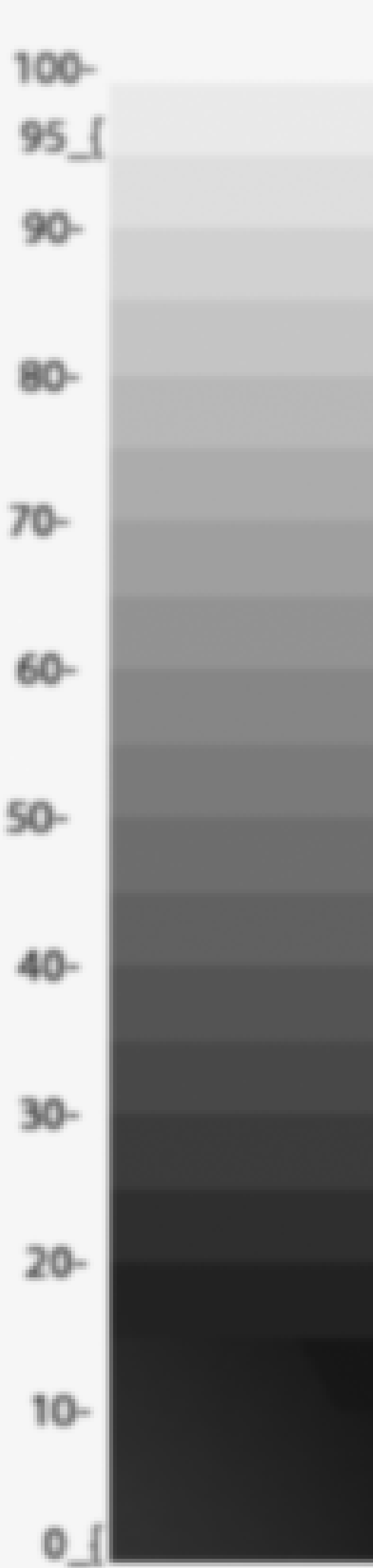}\\
Test Image & After Physical & After Analog\\
\end{tabular}
\caption{Channel contrast changes. Contrast is lost between darker values and gained between lighter ones.
}
\figlabel{fig:effects_contrast}
\end{center}\end{figure}

Our physical channel setup displays each 120-pixel-tall spoof image so the image will fill most of the 480 pixel webcam height.
As a result, each pixel of the spoof image fills approximately a 4x4 grid of
pixels in the image captured by the webcam, though 
  we make no attempt to position the spoof image to align exactly with a grid of four by four webcam pixels and the displayed image does not fill the webcam
  height exactly.
Our analog channel simulates this by upscaling the 120x96 spoof image to a 480x384 image (using pixel replication, i.e., each pixel is replicated in a 4x4 grid) before applying the other effects.

\paragraph{Analog Channel}
Our analog channel simply chains together these individual effects:
we apply upscaling, center brightening,
response curve (contrast reduction), Gaussian blur, and Gaussian noise,
in that order.
Though our \analogchan/ is not a perfect simulation of the physical world's effects, 
   it captures many of the effects that appear to affect face detection
   and has allowed us to get useful results.
Our evaluation of the \analogchan/ is described later (\secref*{sec:analog_eval}).

\section{Oracle}\seclabel{sec:oracle}

\begin{figure}[tbp]\begin{center}
\includegraphics[width=.35\textwidth]{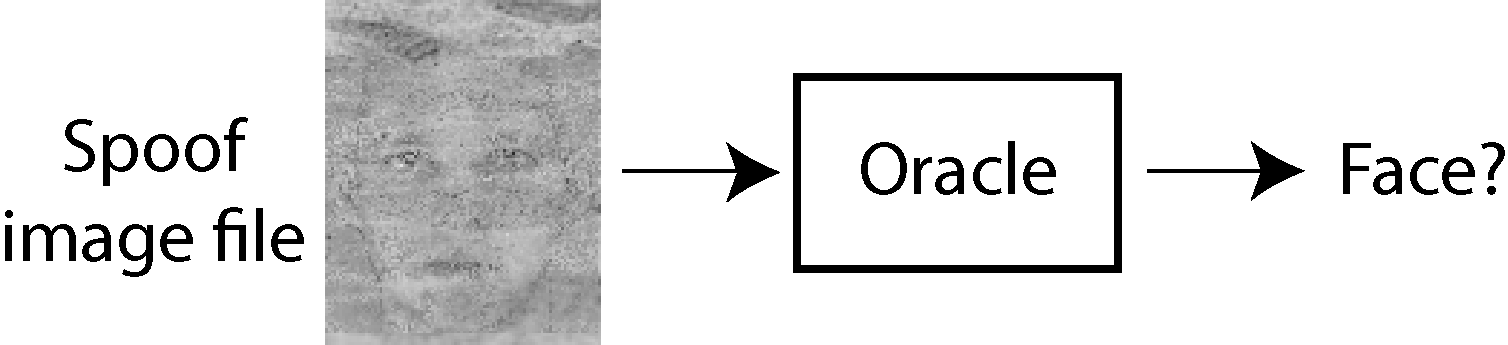}
\caption{A naive oracle for our search algorithm.}
\figlabel{fig:oracle_blackbox}
\end{center}\end{figure}  

We use the analog channel to build an \textit{oracle} 
  (\figref{fig:oracle_blackbox}) which predicts
whether a face will be detected after an image is degraded by the physical
channel.

Naively, we might simply apply the analog channel to the image
and then apply the Viola-Jones to the result.
However, because the noise is partially random, this is not a good
predictor of whether an image will be \textit{reliably} detected as a face:
   even though Viola-Jones detects a face in the degraded image,
   the noise might have just been in our favor that one time.

It is more useful to repeat the procedure multiple times.
Our oracle passes the image in parallel through several copies of the analog channel, 
  runs Viola-Jones on each result, and reports how many times Viola-Jones detected a face (see \figref{fig:oracle_internal}).

\begin{figure}[tbp]\begin{center}
\includegraphics[width=.42\textwidth]{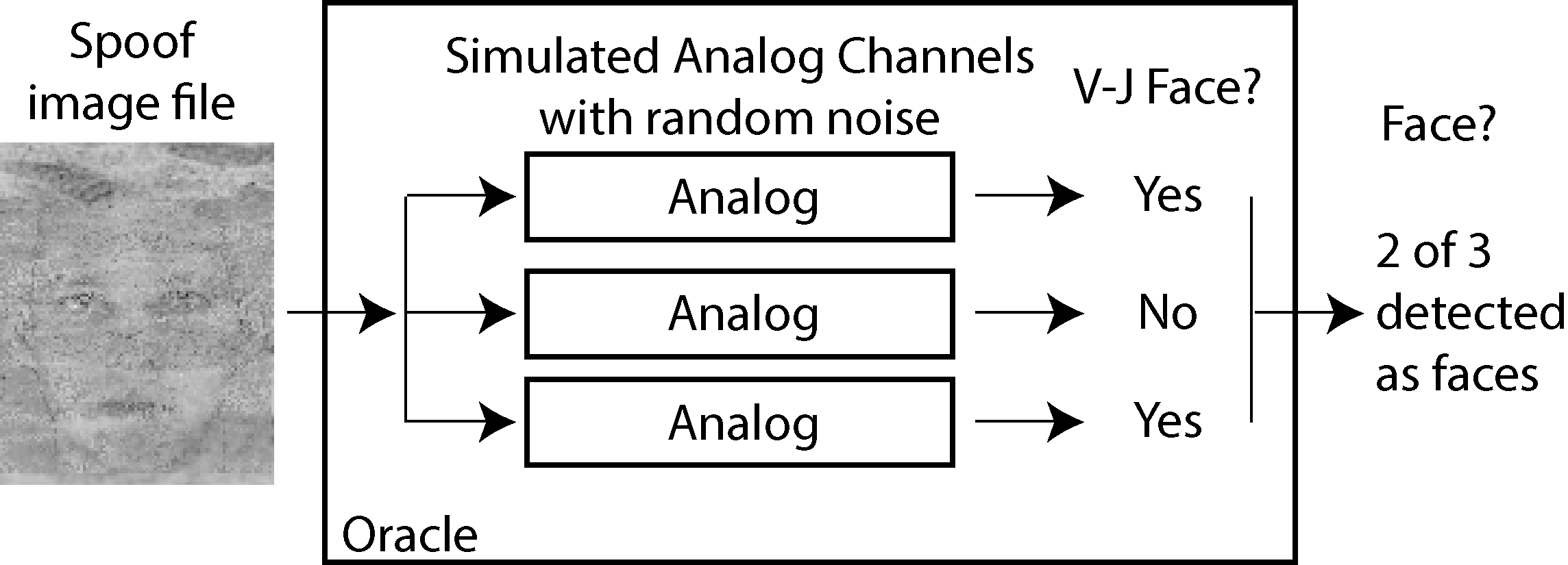}
\caption{Our oracle measures how often a face is detected when repeatedly subjecting the image to the analog channel.}
\figlabel{fig:oracle_internal}
\end{center}\end{figure}

\section{Attack: Random Shift (Analog)}\seclabel{sec:attack_analog}  %
We use our simulation of the physical world to improve the attack
and generate spoof images that are more robust to degradation and
noise from the real world.
We use the same simple search procedure as before:
  pick a random pixel, move its intensity closer to the corresponding pixel in the cover image, and discard the change if our oracle reports failure.
But now we use our upgraded oracle that uses several copies of the analog channel.
Also, for each search, we allow shifting the pixel value not just $50\%$ toward the cover image,
  but some shift rate, $s \in (0, 1)$.

\begin{algorithm}  %

\Fn{Search($F$, $C$, $s$, $n$, $m$)}{
    \Input{Face image $F$,\\
             Cover image $C$, \\
              Shift rate $s$,\\
              Required detections by Oracle $n$,\\
              Number of tests by Oracle $m$
              }
    \Output{Spoof image $S$}

$S \leftarrow F$\;
\While{not Stalled($S$)}{
   $p' \gets $ a random pixel in $S$   %
   
  $\displaystyle
    T(p) \leftarrow 
  \begin{cases}
     \text{Round} \left( S(p) \times (1-s) + C(p) \times s  \right) & \text{if } p = p' \\
     S(p)       & \text{o.w. }
    \end{cases}
  $

  \If{Oracle($T$, $m$)$ \ge n$}{
     $S \leftarrow T$\:
  }
  } %
return S
} %

\Fn{Stalled($S$)}{
    \Input{Image $S$} 
    \Output{Boolean}
    Return True if $S$ has not changed in last several calls.
}

\Fn{Oracle($T$, $m$)}{
    \Input{Test Image $T$} 
    \Output{Boolean}
    Invoke Viola-Jones(Analog($T$)) $m$ times and return how many times it detects a face with bounds within $10\%$ of those of the face in $F$.
}

\caption{Analog attack}
\label{alg:analog}
\end{algorithm}

Initially, we tested the attack with a naive oracle that only applies the analog channel once.
However, we found that the search quickly gets driven towards local minima:
 it tries a change that actually causes the image to be detected only occasionally (say, 10\% of the time), but due to bad luck, the image is accepted by the oracle.
Because the search algorithm tries many candidate changes, many of which are bad, eventually it will get unlucky and accept a bad change.
Once it has moved towards a bad image, the algorithm is unable to recover.
As the search progresses, bad choices vastly outnumber good choices --- most choices are bad --- so the algorithm has a high probability of going awry.

We next tested an oracle that tries multiple times and requires a face be detected at least 2 out of 3 times (or 4 out of 5 times), but we found this suffers from the same problem.
Therefore, we settled on an oracle applies the analog channel 10 times and requires that a face be detected in all 10 out of 10 trials.
We found that this was sufficient to fix the problem.

\paragraph{Experiment}
We prepare the input face the same way as in \secref*{sec:attack_exact}.
We use cover images of granite and all-white and a shift rate of $s=0.7$.
Low shift rates (e.g., $s=0.05$) take a long time and tend to create a faint but visible shadow of a face.
Moderately high shift rates (e.g., $s=0.7$) and an all-white cover image tend to create abstract dot art.
Very high shift rates (e.g., $s=0.9$) tend to fail quickly.
Each run takes a few hours.
We have not yet parallelized the analog channel nor used a GPU.
The analog channel is expensive, though about a third of the cost is Viola-Jones anyway.

\begin{figure}[tbp]
\begin{center}
\begin{tabular}{c@{\hskip3pt}c@{\hskip1pt}c@{\hskip1pt}c@{\hskip1pt}c@{\hskip1pt}c}
iters  & 16,144  & 17,005 & 22,614 & 26,681 \\
\begin{turn}{90}$\quad$ Spoofs\end{turn}&
\includegraphics[width=.20\columnwidth]{%
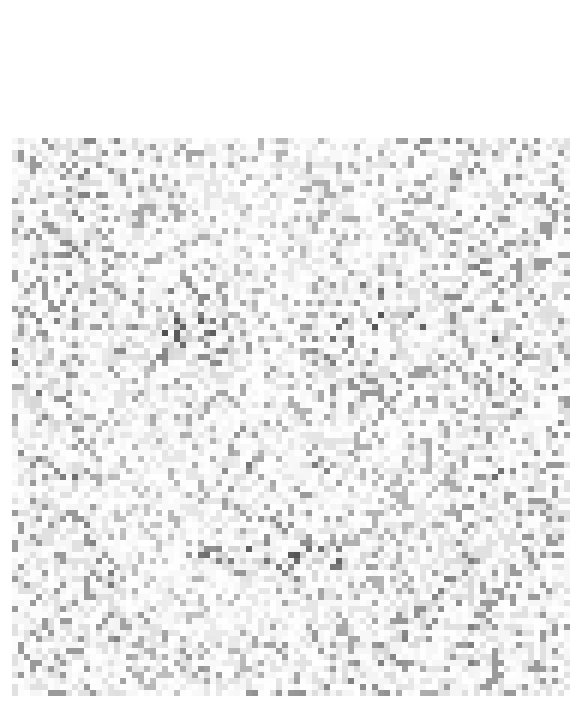}&
\includegraphics[width=.20\columnwidth]{%
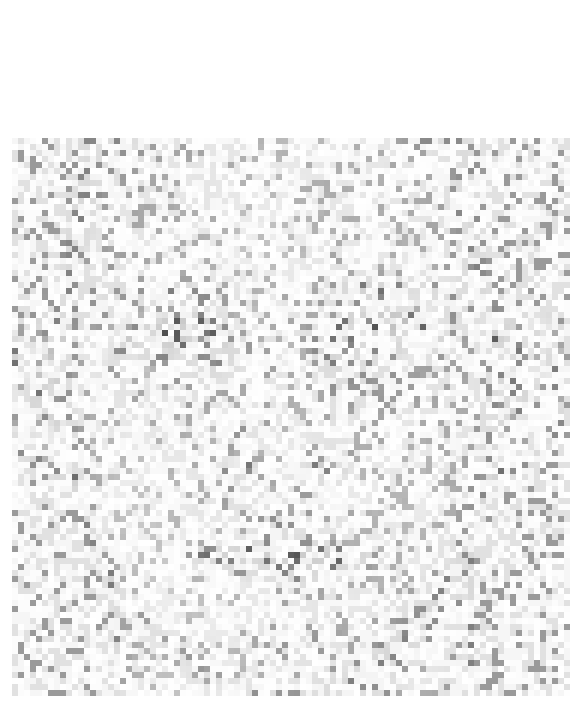}&
\includegraphics[width=.20\columnwidth]{%
figs/20160803_white_10of10_s0p7/exact/big_spoof_0p7_S____22,614_wasY}&
\includegraphics[width=.20\columnwidth]{%
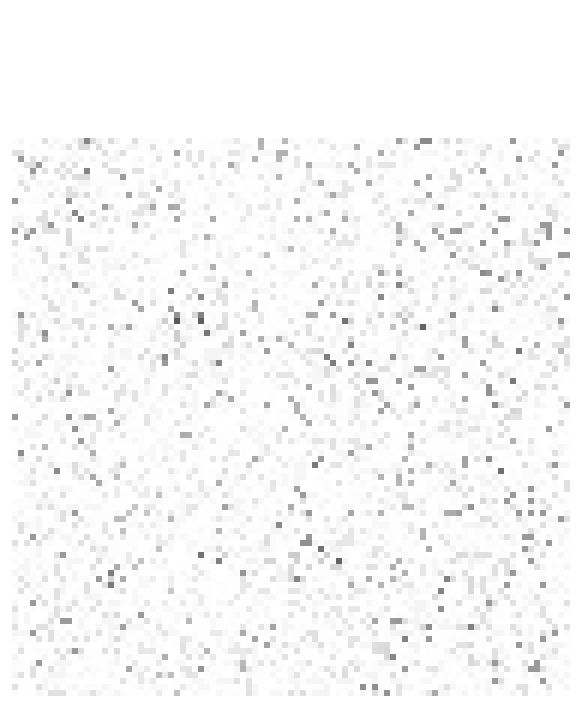}\\
\begin{turn}{90}$\quad$After Analog\end{turn}&
\includegraphics[width=.20\columnwidth]{%
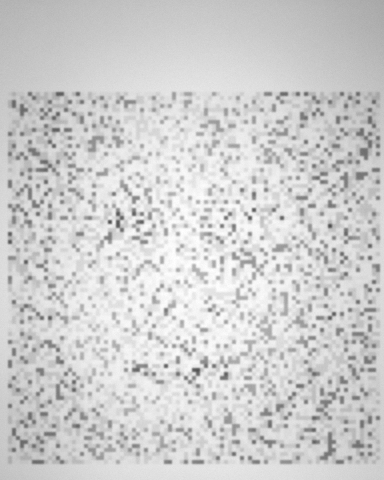}&
\includegraphics[width=.20\columnwidth]{%
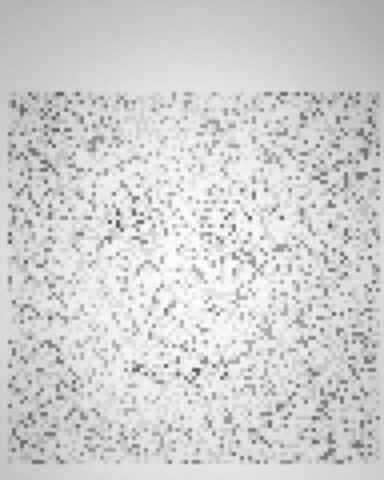}&
\includegraphics[width=.20\columnwidth]{%
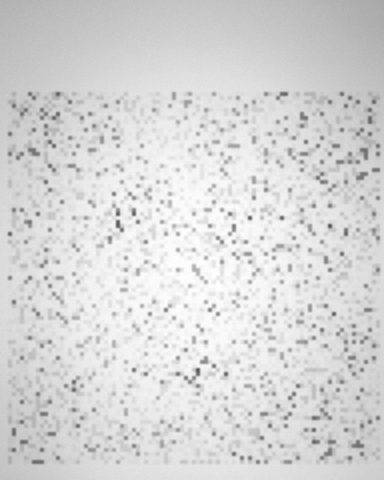}&
\includegraphics[width=.20\columnwidth]{%
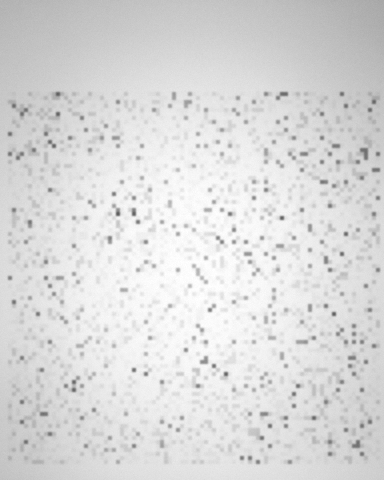}\\
detect \% &   100  &  100  &  87 &  96\\
\begin{turn}{90}$\quad$After Physical\end{turn}&
\includegraphics[width=.20\columnwidth, trim=55mm 0 55mm 0, clip]{%
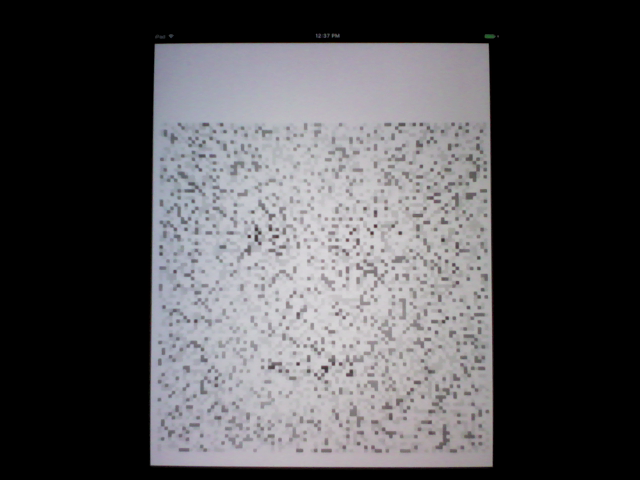}&
\includegraphics[width=.20\columnwidth, trim=55mm 0 55mm 0, clip]{%
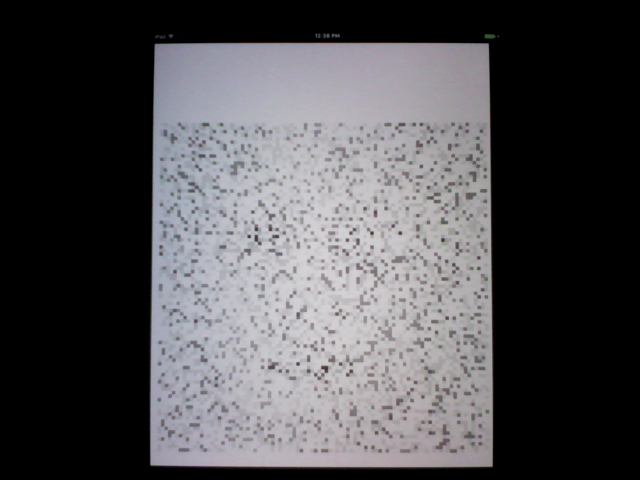}&
\includegraphics[width=.20\columnwidth, trim=55mm 0 55mm 0, clip]{%
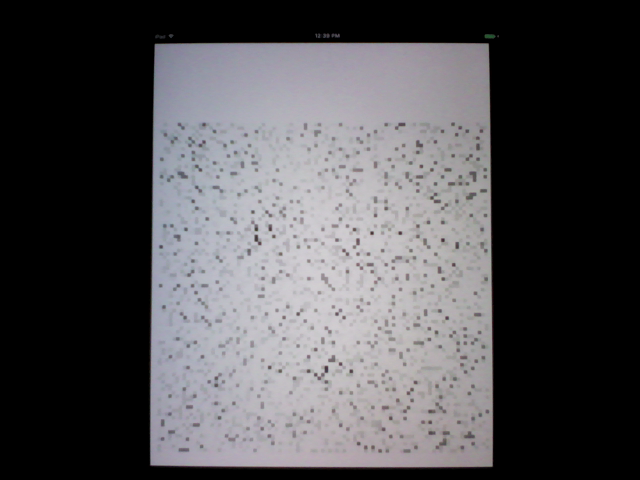}&
\includegraphics[width=.20\columnwidth, trim=55mm 0 55mm 0, clip]{%
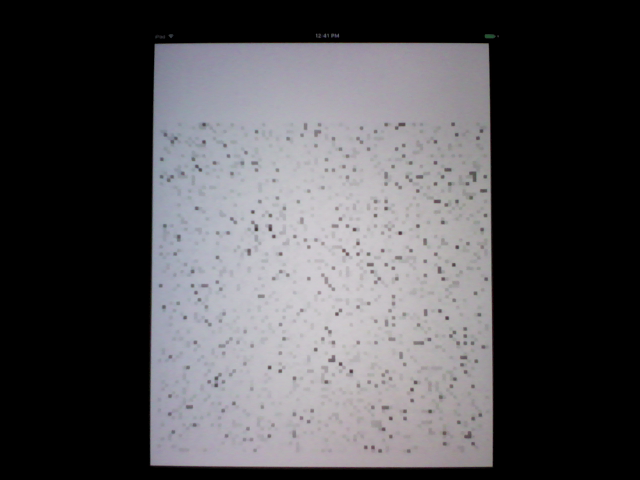}\\
detect \% &   18  &  9 &  97 &  70\\
\end{tabular}
\caption{Spoof images generated by our algorithm, using an all-white cover
image and shift rate $s=0.7$.
We also show an example of applying the analog and physical channel to each
image and the detection rates after each channel.
}
\figlabel{fig:results_analog}
\end{center}
\end{figure}

\paragraph{Results}
\Figref{fig:results_analog} shows a spoof image generated using this algorithm
and an all-white cover image: see the two images in the upper-right.
These images do not stand out to us as faces --- they look rather like some kind of abstract art --- but Viola-Jones detects a face in them, even after applying the physical channel.
The image on the upper-right is detected as a face 70\% of the time
(after the physical channel); the image to its left is detected as a face
97\% of the time, i.e., very reliably.

The four columns in \figref{fig:results_analog} correspond to different points
in time along the evolution of a single run of the search algorithm.
We show the number of iterations so far, the current spoof image (i.e., $S$),
an example of applying our analog channel to that image
and the detection rate after applying the analog channel many times,
as well as an example image after applying the physical channel
and the detection rate after applying the physical channel many times.

Running our algorithm with a granite cover image was not successful.
It takes 9,000 iterations until the algorithm generates a candidate spoof
image that is no longer human-recognizable,
  but the images stop being recognized by Viola-Jones as faces well before that ---
  after 6,000 iterations,
  the detection rate after the physical channel drops to zero.

\paragraph{Discussion}
Our approach is successful at creating images
   that are often detected by Viola-Jones as a faces, 
    but which are not as noticed by humans as faces.
The images are not as stealthy to humans as before,
  but they are more robust: they are detected even after being
  displayed on a tablet and then captured by a camera.

Our attack uses Viola-Jones solely as a black box,
obtaining only a boolean result from it.
One can view the randomized analog channel and 10-out-of-10 oracle as a
way of obtaining a probabilistic measure of success
(a continuous confidence metric) from this black box.
Thus, our techniques might be of independent interest for attacking
other machine learning classifiers, specifically in situations
where the attacker is forced to
use the classifier solely as a black box and cannot obtain any kind
of confidence score, likelihood estimate, or other quantitative measure
from the classifier.

\begin{figure}[tbp]\begin{center}
\includegraphics[width=.45\textwidth]{%
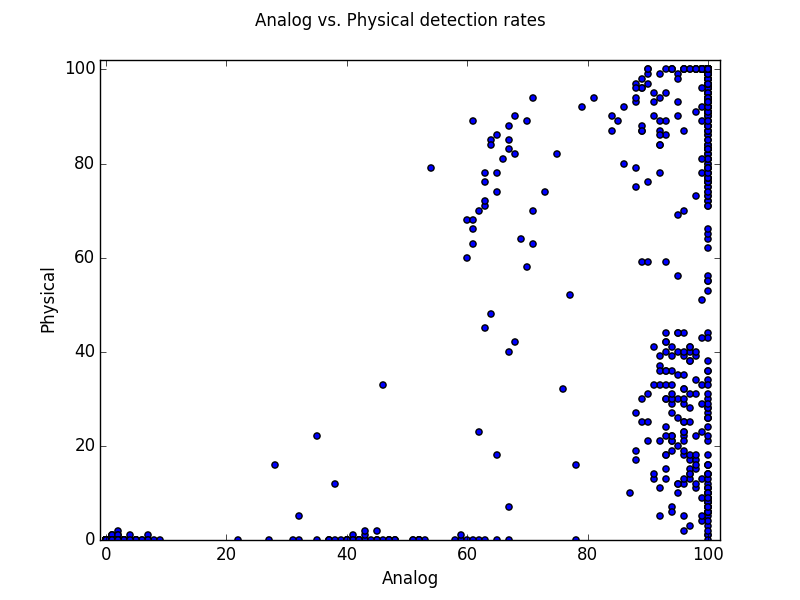}
\caption{Evaluating how well the analog channel models the physical channel.
We plot the face detection rate after the physical channel vs.
face detection rate after the analog channel, for 1200 different images
seen at different iterations of our algorithm.
The analog channel helps avoid some images that would fail the physical channel, but far from guarantees success:
the physical detection rate is generally at most the analog detection rate,
but often substantially smaller.
}
\figlabel{fig:analog_eval}
\end{center}\end{figure}

\section{Evaluation of  Analog Channel}\seclabel{sec:analog_eval}
To evaluate the effectiveness of our simulated \analogchan/,
  we ran many images through both the analog channel and physical channel
  to compare the face detection rate after each.
We gathered 1200 images seen during runs of our algorithm.
For each image, we fed it through the physical channel 100 times and
counted how many times Viola-Jones detected a face in the result.
We also did the same for the analog channel.
\Figref{fig:analog_eval} shows a scatterplot of the resulting scores.
We see that the score from the analog channel is an imperfect but
useful predictor of the physical channel: the analog channel helps us
rule out some images that won't survive the physical channel, but is
sometimes too optimistic about the likelihood that Viola-Jones will detect
a face after the physical channel is applied.
This explains why our oracle helps improve the search algorithm:
while not perfect,
it provides feedback to help the random search avoid some images
that certainly won't survive the physical channel.

\section{Failed Attack: Gradient Descent}\seclabel{sec:attack_delta}
Before arriving at the simple algorithm described earlier, we tried
other approaches.
Most notably, we tried an approach inspired by gradient descent, where we
tried to measure which pixels Viola-Jones is most sensitive to.

To approximate the gradient of the detector's confidence that the
image is a face, we first found a blend of the current
image and the cover image that was right on the edge of Viola-Jones'
decision boundary.
In particular, we used binary search to find 
the largest real number $\alpha \in [0,1]$ such that
the blend $(1-\alpha) \times S + \alpha \times C$ was still detected
by Viola-Jones, then reduced $\alpha$ slightly and set
$S' = (1-\alpha) \times S + \alpha \times C$.
This gave us an image $S'$ in the search space close to the decision boundary.
We then picked small regions of pixels and saw how close to the cover image we could move them as a group
  before Viola-Jones failed. 
Doing this for overlapping regions gave us a measure of how sensitive Viola-Jones was to changes to each pixel.
In this way, we built an approximate sensitivity map $M$, where
$M(p)$ is proportional to how far we can move pixel $p$ of $T$ towards $C(p)$
before Viola-Jones stops detecting a face
(i.e., larger values of $M(p)$ indicate lower sensitivity to changes
to $p$).
Then, we used this information to move the original image $S$
closer to the cover image, weighted to move the less sensitive pixels the most:
i.e., we replaced $S$ with the image $T$ defined by
$$T(p) = (1-\varepsilon \cdot M(p)) \times S(p) + \varepsilon \cdot M(p) \times C(p),$$
for some small constant $\varepsilon>0$.
We iterated this process until convergence.

While slow, this process created images with just the key face features left.
Unfortunately, those images were readily human-recognizable as faces;
see, e.g., \figref{fig:attack_delta}.
The randomness of our current approach seems to hide the face better.

\begin{figure}[htbp]
\begin{center}
\includegraphics[width=.2\textwidth]{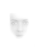}
\caption{An image generated using our gradient descent-inspired attack.}
\figlabel{fig:attack_delta}
\end{center}\end{figure}

\section{Related Work}\seclabel{sec:related}
There has not been published work on stealthy spoofs of 2D face detection.
Work in the opposite direction has tried to hide from facial detection 
  with obvious\cite{harvey_cvdazzle} or less obvious\cite{eckert2013facial} makeup or partial obscurement. 

Recent work on physical world attacks on object recognition
measured the degradation in effectiveness of adversarial images
when they are first printed and photographed with a cell phone\cite{kurakin_adversarial_2016}.
Though they measure the effects of components of that physical channel,
  they construct their images using knowledge of the object detection algorithm, and not the channel.
Our work constructs adversarial images without knowledge of the detection algorithm.

\section{Conclusion}
We have shown that deliberate spoof images can be created 
  that do not appear to humans as faces,
  yet Viola-Jones often detects as faces, even after passing through a simulated physical world.
This indicates that facial detection can be fooled, and in a way that human observers are unlikely to notice as suspicious.

\bibliographystyle{abbrv}
\bibliography{2DSpoof.bib}

\end{document}